\documentclass[aps,preprintnumbers,nofootinbibt,referee, superscriptaddress, keywords]{revtex4}
\usepackage{graphicx}
\usepackage{amsmath}
\usepackage{bm}
\usepackage{color}

\def\be{\begin{equation}}
\def\ee{\end{equation}}
\def\bea{\begin{eqnarray}}
\def\eea{\end{eqnarray}}

\begin{document}

\title{A simple computational approach to the Susceptible-Infected-Recovered (SIR)
epidemic model via the Laplace-Adomian Decomposition Method}
\author{Tiberiu Harko}
\email{tiberiu.harko@aira.astro.ro}
\affiliation{Astronomical Observatory, 19 Ciresilor Street, 400487 Cluj-Napoca, Romania,}
\affiliation{Faculty of Physics, Babes-Bolyai University, 1 Kogalniceanu Street,
400084 Cluj-Napoca, Romania}
\affiliation{School of Physics, Sun Yat-Sen University,  Xingang  Road, 510275 Guangzhou, People's
Republic of China}
\author{Man Kwong Mak}
\affiliation{Departamento de Fisica, Facultad de Ciencias Naturales, Universidad de Atacama, Copayapu 485,
Copiapo, Chile}
\email{mankwongmak@gmail.com}

\begin{abstract}
The Susceptible-Infected-Recovered (SIR) epidemic model is extensively used for the study of the spread of infectious diseases. Even that the exact solution of the model can be obtained in an exact parametric form, in order to perform the comparison with the epidemiological data a simple but highly accurate representation of the time evolution of the SIR compartments would be very useful. In the present paper we obtain a series representation of the solution of the SIR model by using the Laplace-Adomian Decomposition Method to solve the basic evolution equation of the model. The solutions  are expressed in the form of infinite series. The series representations of the time evolution of the SIR compartments are compared with the exact numerical solutions of the model. We find that there is a good agreement between the Laplace-Adomian semianalytical solutions containing only three terms, and the numerical results.

\end{abstract}

\maketitle

\tableofcontents


\section{Introduction}

The major Covid-19 coronavirus pandemic that began in 2019 in Wuhan, China \cite{China}, has again underlined the necessity of understanding on both quantitative and qualitative levels the outbreak and spread of infectious diseases. This problem has been already studied a long time ago,  John Graunt being the first scientist who attempted to
evaluate in a methodical way the causes of fatalities during epidemics \cite{1}. His analysis led to a theory that is still widely accepted by the modern
epidemiologists. The first mathematician who did propose a mathematical model describing the dynamics of an infectious disease was Daniel Bernoulli, who  modeled the spread of Variola,  which was rampant at the time, in 1760 \cite{Ber1}. Moreover, in a later study Daniel Bernoulli did present
the advantages of inoculation \cite{Ber2},  the method first used to immunize an individual against smallpox  with material taken from a patient.

More recently an important simple (compartmental) deterministic
model predicting the spread of epidemic outbreaks was proposed  by McKendrick and Kermack in 1927 \cite{2}. This
mathematical epidemic model is called the Susceptible-Infected-Recovered (SIR)
model, or the $xyz$ model. In the SIR model it is assumed that in the presence of an infectious disease a population with a fixed number can be compartmentalized into three groups (compartments), denoted (S), (I) and (R), respectively. More precisely, the compartments used in the McKendrick-Kermack model  are defined as follows \cite{2}:

a) The (S) (susceptible) compartment consists of the individuals in the total population not yet infected
with the disease at time $t$, or those susceptible to the disease. The number of individuals in this compartment is denoted $x(t)$.

b) The (I) (infected) compartment consists of the individuals who have been
infected with the epidemic disease, and who can spread the disease to those
in the susceptible category. The number of individuals in the (I) compartment is denoted $y\left( t\right) $.

c) The (R) (recovered) compartment consists of the individuals who have been
infected during the epidemics, and did fully recover.  The individuals in the (R) compartment, whose number is denoted by $z(t)$,  cannot be infected again, and they are not able to transmit the disease to others.

From a mathematical point of view the SIR model is formulated in terms of a strongly nonlinear system of ordinary differential
equations. The behavior of the SIR model essentially depends on two constants $\beta $ and $\gamma $ that give the transition rates between
compartments.

The transition rate between the S (Susceptible) and I (infected) groups is
denoted by $\beta $.  It represents the contact rate, that is,
the probability of being infected when the contact between a susceptible
and an infectious individual takes place \cite{14,Mur, 15, Mur1}.

The transition rate
between the I (Infected) and R (recovered) compartments is denoted by $\gamma $.  $\gamma $ gives
the rate of recovery or death. If we denote by $D$ the total duration of the infection, then $\gamma = 1/D$, expressing the fact that an individual experiences one recovery in $D$ units of time \cite{14,Mur, 15, Mur1}. Since in the SIR model both $\beta $ and $%
\gamma $ represent transition rates (probabilities), their range of values is $0\leq \beta \leq 1$ and $0\leq \gamma \leq 1$, respectively.

The SIR model has been investigated from both mathematical and epidemiological point of view in a large number of studies. Its exact solution has been obtained in a parametric form in \cite{Harko1}. The SIR model has been used extensively for the simulation of the dynamics of the Covid-19 pandemic and its impact on society \cite{Cov1, Cov2,Cov3,Cov4,Cov5,Cov6,Cov7,Cov8,Cov9,Cov10,Cov11}. Despite the large number of mathematical and numerical investigations of the SIR model, including the use of the Adomian decomposition method \cite{5}, of the variational iteration method \cite{6}, of the homotopy
perturbation method \cite{7}, and of the differential transformation method \cite{8}, a simple but precise representation of the solution of the SIR model that could be efficiently used in epidemiological and populations studies is still missing.

It is the purpose of this work to present a simple series solution of the SIR epidemic model. To achieve this goal we first derive the basic equation of the SIR model, which is represented by a first order nonlinear differential equation for $z(t)$. Even that this equation can be solved exactly, in a parametric form, we will investigate it by using the Laplace-Adomian Decomposition Method (LADM), which is an extension of the standard Adomian Decomposition Method \cite{Ad1,Ad2,Ad3}. The ADM is
a powerful mathematical method that permit to find semianalytical solutions of different types of ordinary, partial
and integral differential nonlinear equations. The ADM allows to obtain the solution of a differential equation in the form of a series, with the terms of the series determined recursively by using the Adomian polynomials \cite{Ad1,Ad2}.  The Laplace-Adomian Method has been used to investigate nonlinear differential equations in different fields of science in \cite{Ad4,Ad5,Ad6, Ad7,Ad8,Ad9, Ad10}. By applying the Laplace-Adomian Method to the basic equation of the SIR model its solution can be represented as a series that gives an excellent description of the numerical results.

The present paper is organized as follows. The series solution of the SIR
epidemic model is obtained in Section~\ref{II}, where its comparison with the exact numerical solution is also performed.  We briefly conclude our
results in Section~\ref{IV}.

\section{ Series solution of the SIR epidemic model}\label{II}

In the present Section we will present a semianalytical solution of the SIR model represented in the form of a series containing exponential terms. The basic equations of the SIR model are
\begin{equation}
\frac{dx}{dt}=-\beta x\left( t\right) y\left( t\right) ,  \label{A1}
\end{equation}%
\begin{equation}
\frac{dy}{dt}=\beta x\left( t\right) y\left( t\right) -\gamma y\left(
t\right) ,  \label{A2}
\end{equation}%
and
\begin{equation}
\frac{dz}{dt}=\gamma y\left( t\right) ,  \label{A3}
\end{equation}%
respectively, where $x(t)>0$, $y(t)>0$ and $z(t)>0$, $\forall t\geq 0$. The system of equations (\ref{A1})-(\ref{A3}) must be integrated with the initial conditions $x\left( 0\right) =N_{1}\geq 0$, $%
y\left( 0\right) =N_{2}\geq 0$ and $z\left( 0\right) =N_{3}\geq 0$, respectively, where $%
N_{i}\in \Re $, $i=1,2,3$. The time evolution of the model is determined by two epidemiological parameters, the infection rate $\beta $, and the mean recovery rate $\gamma $, which in the following are assumed to be positive constants.

By adding  Eqs.~(\ref{A1})--(\ref{A3}) we immediately obtain the differential equation,
\begin{equation}
\frac{d}{dt}\left[ x(t)+y(t)+z(t)\right] =0,  \label{A4}
\end{equation}%
which upon integration gives the first integral of the SIR system
\begin{equation}
x(t)+y(t)+z(t)=N,\forall t\geq 0,  \label{A5}
\end{equation}%
 Hence it follows that in this epidemiological model the total size of the population $N=\sum _{i=1}^3{N_i}$ is, from a mathematical point of view,  an arbitrary positive integration constant. This result is consistent with our assumption of considering only three compartments in a fixed population with
$N$ members.

\subsection{The basic evolution equation for the SIR model}

We will now show that the SIR model can be equivalently formulated in terms of a first order nonlinear differential equation. After differentiating  Eq.~(\ref{A1}) with respect to the time $t$ we obtain the following second order differential equation,
\begin{equation}
\frac{dy}{dt}=-\frac{1}{\beta }\left[ \frac{x^{\prime \prime }}{x}-\left(
\frac{x^{\prime }}{x}\right) ^{2}\right] ,  \label{A6}
\end{equation}
where in the following we denote by a prime the derivative with respect to time $t$. By
inserting Eqs.~(\ref{A1}) and (\ref{A6}) into Eq.~(\ref{A2}), we obtain
\begin{equation}\label{A7}
\frac{x^{\prime \prime }}{x}-\left( \frac{x^{\prime }}{x}\right) ^{2}+\gamma
\frac{x^{\prime }}{x}-\beta x^{\prime }=0.
\end{equation}

From Eqs. (\ref{A1}) and (\ref{A3}) we find the evolution equation for $z$, in terms of $x$ as
\begin{equation}
\frac{dz}{dt}=-\frac{\gamma }{\beta }\left( \frac{x^{\prime }}{x}\right) \,.
\label{A8}
\end{equation}
which  can be integrated to give
\begin{equation}
x=x_{0}e^{-(\beta /\gamma )z},  \label{A82}
\end{equation}
The value of the positive integration constant $x_{0}$ can be obtained from the initial conditions of the SIR model as
\begin{equation}
x_0=N_1e^{(\beta /\gamma )N_3}.
\end{equation}

With the use of Eq.~(\ref{A7}),  Eq.~(\ref{A6}) becomes
\be
\frac{dy}{dt}=-x'+\frac{\gamma }{\beta}\frac{x'}{x},
\ee
immediately giving
\be
y(t)=\frac{\gamma }{\beta}\ln x-x+C_1,
\ee
where $C_1$ is an arbitrary integration constant, which can be determined from the initial conditions as
\be
C_1=N_1+N_2-\frac{\gamma  }{\beta}\ln N_1.
\ee

After differentiating  Eqs.~(\ref{A82}) and (\ref{A8}) with respect to the time $t$, we obtain the basic second order differential equation for $z(t)$, describing  the spread of the non-fatal disease in a given population,
\begin{equation}\label{A10a}
z^{\prime \prime }=x_{0}\beta z^{\prime }e^{-\frac{\beta }{\gamma }z}-\gamma
z^{\prime }.
\end{equation}

Eq.~(\ref{A10a}), a strongly nonlinear differential equation,  is mathematically equivalent to the first order system of differential equations Eqs.~(%
\ref{A1})--(\ref{A3}), respectively, giving the SIR epidemic model. However, a first order nonlinear differential equation can also de obtained to describe the evolution of $z$.  Eq.~(\ref{A10a}) can be rewritten as
\be
\frac{d^2z}{dt^2}+\gamma \frac{dz}{dt}=-\gamma x_0\frac{d}{dt}e^{-(\beta/\gamma)z},
\ee
and thus we obtain its first integral as given by
\be\label{Eqzf}
\frac{dz}{dt}+\gamma z=-\gamma x_0e^{-(\beta/\gamma)z}+C,
\ee
where $C$ is an arbitrary constant of integration. By estimating the above equation at $t=0$ and with the help of Eq.~(\ref{A3}) we obtain the value of the integration constant as
\be
C=\gamma \left(N_2+N_3\right)+\gamma x_0e^{-\frac{\beta }{\gamma }N_3}=\gamma N.
\ee

\subsection{Series solution of the basic equation of the SIR model via the Laplace-Adomian method}

The general solution of Eq.~(\ref{Eqzf}) can be obtained in a closed form, in an exact parametric representation. However, the parametric representation is not particularly useful from a computational point of view. That's why in the following we will obtain a series solution of Eq.~(\ref{Eqzf}). In order to do this we will use the Laplace-Adomian method.

In the Laplace-Adomian method we first apply the Laplace transformation
operator $\mathcal{L}_x$, defined as  $\mathcal{L}_x[f(x)](s) = \int_0^{\infty}{f(x)e^{-sx}dx}$, to Eq.~(\ref{Eqzf}), thus obtaining
\begin{equation}
\mathcal{L}_t\left[\frac{dz(t)}{dt}\right](s) +\gamma \mathcal{L}_t[z(t)](s)-
\mathcal{L}_t[C](s)=-\gamma x_0\mathcal{L}_t\left[ e^{-(\beta/\gamma)z(t)}\right](s).
\end{equation}

By using the properties of the Laplace transform we immediately  find
\begin{equation}
(\gamma +s) \mathcal{L}_t[z(t)](s)-\frac{\gamma N+s z(0)}{s} =-\gamma x_0\mathcal{L}_t\left[ e^{-(\beta/\gamma)z(t)}\right](s).
\end{equation}

With the use of the initial condition $z(0)=N_3$ we obtain
\begin{equation}\label{6a}
\mathcal{L}_t[z(t)](s)=\frac{\gamma N+N_3s}{s(\gamma +s)}-\frac{\gamma x_0}{\gamma +s}\mathcal{L}_t\left[ e^{-(\beta/\gamma)z(t)}\right](s).
\end{equation}
We assume now that the function $z(t)$ can be represented in the form of an infinite series,
\begin{equation}\label{7a}
z(t)=\sum_{n=0}^{\infty }z_{n}(t),
\end{equation}%
where the terms $z_{n}(t)$ can be computed recursively. As for the nonlinear
operator  $f(z(t))=-\gamma x_0e^{-(\beta/\gamma)z(t)}$, we assume that it can be decomposed as
\begin{equation}\label{8a}
f(z(t))=-\gamma x_0e^{-(\beta/\gamma)z(t)}=\sum_{n=0}^{\infty }A_{n}(t),
\end{equation}%
where the $A_{n}$'s are Adomian polynomials, defined generally for an arbitrary function $f(z(t))$
according to \cite{Ad3}
\begin{equation}
A_{n}=\left. \frac{1}{n!}\frac{d^{n}}{d\lambda ^{n}}f\left(
\sum_{i=0}^{\infty }{\lambda ^{i}z_{i}}\right) \right\vert _{\lambda =0}.
\end{equation}

The first five Adomian polynomials can be obtained in the following form,
\begin{equation}
A_{0}=f\left( z_0\right) ,  \label{Ad0}
\end{equation}%
\begin{equation}
A_{1}=z_{1}f^{\prime }\left( z_0\right) ,  \label{Ad1}
\end{equation}%
\begin{equation}
A_{2}=z_{2}f^{\prime }\left( z_0\right) +\frac{1}{2}z_{1}^{2}f^{\prime
\prime }\left( z_0\right) ,  \label{Ad2}
\end{equation}%
\begin{equation}
A_{3}=z_{3}f^{\prime }\left( z_0\right) +z_{1}z_{2}f^{\prime \prime }\left(
z_0\right) +\frac{1}{6}z_{1}^{3}f^{\prime \prime \prime }\left( z_0\right) ,
\label{Ad3}
\end{equation}
\begin{equation}
A_{4}=z_{4}f^{\prime }\left( z_0\right) +\left[ \frac{1}{2!}%
z_{2}^{2}+z_{1}z_{3}\right] f^{\prime \prime }\left( z_0\right) +\frac{1}{2!}%
z_{1}^{2}z_{2}f^{\prime \prime \prime }\left( z_0\right) +\frac{1}{4!}%
z_{1}^{4}f^{(\mathrm{iv})}\left( z_0\right) .  \label{Ad4}
\end{equation}

Substituting Eqs. (\ref{7a}) and (\ref{8a}) into Eq. (\ref{6a}) we obtain
\begin{equation}
\mathcal{L}_t\left[ \sum_{n=0}^{\infty }z_{n}(t)\right] =\frac{\gamma N+N_3s}{s(\gamma +s)}-\frac{1}{%
\gamma+s}\mathcal{L}_t[\sum_{n=0}^{\infty }A_{n}].  \label{11a}
\end{equation}

The matching of both sides of Eq. (\ref{11a}) gives the following iterative
algorithm for the power series solution of the basic evolution equation of the SIR model, Eq.~(\ref{Eqzf}),
\begin{equation}\label{12a}
\mathcal{L}_t\left[ z_{0}(t)\right](s) =\frac{\gamma N+N_3s}{s(\gamma +s)},
\end{equation}%
\begin{equation}\label{12b}
\mathcal{L}_t\left[ z_{1}(t)\right](s) =-\frac{1}{%
\gamma+s}\mathcal{L}_t\left[A_{0}(t)\right](s) ,
\end{equation}%
\begin{equation}
\mathcal{L}_t\left[ z_{2}(t)\right](s) =-\frac{1}{s+\gamma }\mathcal{L}_t\left[
A_{1}(t)\right](s) ,  \label{12c}
\end{equation}%
\begin{equation*}
...
\end{equation*}%
\begin{equation}
\mathcal{L}_t\left[ z_{k+1}(t)\right](s) =-\frac{1}{s+\gamma}\mathcal{L}_t%
\left[ A_{k}(t)\right](s) .  \label{12n}
\end{equation}

By applying the inverse Laplace transformation to Eq. (\ref{12a}), we obtain
the value of $z_{0}$ as
\be
z_0(t)=\mathcal{L}_t^{-1}\left\{\left[\frac{\gamma N+N_3s}{s(\gamma +s)}\right](s)\right\}(t)=N-\left(N_1+N_2\right) e^{-\gamma t}.
\ee
 In the following, for computational simplicity, we will approximate $z_0(t)$ by its first order series expansion, given by
 \be
 z_0(t)\approx N_3+\gamma \left(N_1+N_2\right)t.
 \ee

The first Adomian polynomial is given by $A_0(t)=-\gamma x_0e^{-(\beta/\gamma)z_0(t)}$, and thus, within the adopted approximation, we obtain
\be
A_0(t)=-\gamma N_1e^{-\beta \left(N_1+N_2\right)t},
\ee
and
\be
z_1(t)= -\mathcal{L}_t^{-1}\left\{\frac{1}{s+\gamma}\mathcal{L}_t\left[ \gamma N_1e^{-\beta \left(N_1+N_2\right)t}\right](s)\right\}(t)=\frac{\gamma  N_1 \left[e^{-\beta  \left(N_1+N_2\right)t}-e^{-\gamma
   t}\right]}{\beta  \left(N_1+N_2\right)-\gamma },
\ee
respectively. The second Adomian polynomial is obtained as
\be
A_1(t)=-\frac{\beta  \gamma  N_1^2 e^{-\left[\gamma +2 \beta  \left(N_1+N_2\right)\right]t}
   \left[e^{\beta   \left(N_1+N_2\right)t}-e^{\gamma  t}\right]}{\beta
   \left(N_1+N_2\right)-\gamma },
\ee
and thus we find
\bea
\hspace{-0.9cm}&&z_2(t)= \mathcal{L}_t^{-1}\left\{\frac{1}{s+\gamma}\mathcal{L}_t\left[ A_1(t)\right](s)\right\}(t)=\nonumber\\
\hspace{-0.9cm}&&\frac{\gamma  N_1^2 e^{-\left[\gamma +2 \beta  \left(N_1+N_2\right)\right]t} \left\{-\beta
   \left(N_1+N_2\right) e^{\gamma  t}+\left[2 \beta  \left(N_1+N_2\right)-\gamma
   \right] e^{\beta   \left(N_1+N_2\right)t}+\left[\gamma -\beta  \left(N_1+N_2\right)\right] e^{2 \beta
    \left(N_1+N_2\right)t}\right\}}{\left(N_1+N_2\right) \left[\beta
   \left(N_1+N_2\right)-\gamma \right] \left[2 \beta  \left(N_1+N_2\right)-\gamma \right]}.
\eea
Similarly, after computing the third Adomian polynomial, we obtain
\begin{eqnarray}
z_{3}(t) &=&\frac{\gamma N_{1}^{3}}{2\left[ \gamma -\beta \left(
N_{1}+N_{2}\right) \right] ^{2}}\Bigg\{\frac{\beta ^{2}\left[ 4\beta \left(
N_{1}+N_{2}\right) -3\gamma \right] e^{-3\beta \left( N_{1}+N_{2}\right) t}}{%
\gamma ^{2}+6\beta ^{2}\left( N_{1}+N_{2}\right) ^{2}-5\beta \gamma \left(
N_{1}+N_{2}\right) }+\frac{\beta ^{2}e^{-\left[ 2\gamma +\beta \left(
N_{1}+N_{2}\right) \right] t}}{\gamma +\beta \left( N_{1}+N_{2}\right) }+
\nonumber \\
&&\frac{2\left[ \gamma -\beta \left( N_{1}+N_{2}\right) \right] ^{2}e^{-%
\left[ \gamma +\beta \left( N_{1}+N_{2}\right) \right] t}}{\left(
N_{1}+N_{2}\right) ^{2}\left[ 2\beta \left( N_{1}+N_{2}\right) -\gamma %
\right] }+\frac{\left[ \gamma -2\beta \left( N_{1}+N_{2}\right) \right] e^{-%
\left[ \gamma +2\beta \left( N_{1}+N_{2}\right) \right] t}}{\left(
N_{1}+N_{2}\right) ^{2}}-  \nonumber \\
&&\frac{e^{-\gamma t}\left[ \gamma -\beta \left( N_{1}+N_{2}\right) \right]
^{2}\left[ \gamma +2\beta (N_{1}+N_{2})\right] }{\left( N_{1}+N_{2}\right)
^{2}\left[ 3\beta \left( N_{1}+N_{2}\right) -\gamma \right] \left[ \gamma
+\beta \left( N_{1}+N_{2}\right) \right] }\Bigg\}.
\end{eqnarray}

Finally, we will present the $z_4(t)$ term in the Adomian series expansion, which is given by
\begin{eqnarray}
z_4(t)&=&\frac{1}{6} \gamma  N_1^4 \Bigg\{\frac{2 \beta ^3 \left[8 \gamma -9 \beta
   \left(N_1+N_2\right)\right] e^{-4 \beta  \left(N_1+N_2\right)t}}{\left[\beta
   \left(N_1+N_2\right)-\gamma \right]^3 \left[3 \beta  \left(N_1+N_2\right)-\gamma \right] \left[4 \beta
   \left(N_1+N_2\right)-\gamma \right]}+\nonumber\\
  && \frac{\beta ^3 e^{- \left[3 \gamma +\beta
   \left(N_1+N_2\right)\right]t}}{\left[\beta  \left(N_1+N_2\right)-\gamma \right]^3 \left[2 \gamma +\beta
   \left(N_1+N_2\right)\right]}+\nonumber\\
  && \frac{6 \beta ^2 \left[\gamma ^2-2 \beta ^2
   \left(N_1+N_2\right)^2\right] e^{-2  \left[\gamma +\beta
   \left(N_1+N_2\right)\right]t}}{\left(N_1+N_2\right) \left[\beta  \left(N_1+N_2\right)-\gamma \right]^3
   \left[\gamma +\beta  \left(N_1+N_2\right)\right] \left[\gamma +2 \beta  \left(N_1+N_2\right)\right]}+\nonumber\\
  && \frac{6\beta ^{2}e^{-\left[2\gamma +\beta \left(N_1+N_2\right)\right]t}}{\left(N_1+%
N_2\right)\left[\beta \left(N_1+N_2\right)-\gamma \right]\left[2\beta \left(N_1+N_2%
\right)-\gamma \right]\left[\gamma +\beta \left(N_1+N_2\right)\right]}+\nonumber\\
   && \frac{e^{-\gamma t} \left[-2 \gamma ^3-12 \beta ^3
   \left(N_1+N_2\right)^3-28 \beta ^2 \gamma  \left(N_1+N_2\right)^2-13 \beta  \gamma ^2
   \left(N_1+N_2\right)\right]}{\left(N_1+N_2\right)^3 \left[4 \beta
   \left(N_1+N_2\right)-\gamma \right] \left[\gamma +\beta  \left(N_1+N_2\right)\right] \left[\gamma +2 \beta
   \left(N_1+N_2\right)\right] \left[2 \gamma +\beta  \left(N_1+N_2\right)\right]}+\nonumber\\
  && \frac{\left[-\gamma
   ^3+12 \beta ^3 \left(N_1+N_2\right)^3-17 \beta ^2 \gamma  \left(N_1+N_2\right)^2+7
   \beta  \gamma ^2 \left(N_1+N_2\right)\right] e^{- \left[\gamma +3 \beta
   \left(N_1+N_2\right)\right]t}}{\left(N_1+N_2\right)^3 \left[\beta  \left(N_1+N_2\right)-\gamma
   \right]^3 \left[2 \beta  \left(N_1+N_2\right)-\gamma \right]}+\nonumber\\
  && \frac{\left[3 \gamma +6 \beta
   \left(N_1+N_2\right)\right] e^{- \left[\gamma +\beta
   \left(N_1+N_2\right)\right]t}}{\left(N_1+N_2\right)^3 \left[3 \beta  \left(N_1+N_2\right)-\gamma
   \right] \left[\gamma +\beta  \left(N_1+N_2\right)\right]}-\frac{3 e^{- \left[\gamma +2 \beta
   \left(N_1+N_2\right)\right]t}}{\left(N_1+N_2\right)^3 \left[\beta  \left(N_1+N_2\right)-\gamma
   \right]}\Bigg\}.
\end{eqnarray}

The higher order terms in the Laplace-Adomian power series representation of $z(t)$ can be easily obtained with the help of some symbolic computation software. Hence, the solution of the SIR system can be obtained as
\be
x(t)=x_0e^{-\frac{\beta}{\gamma}\sum _{i=0}^{\infty}{z_i(t)}}=N_1e^{(\beta /\gamma)\left(N_3-\sum_{i=0}^{\infty}{z_i(t)}\right)},
\ee
\bea
y(t)&=&N_1+N_2+\frac{\gamma }{\beta}\ln \frac{x_0e^{-\frac{\beta}{\gamma}\sum _{i=0}^n{z_i(t)}}}{N_1}-x_0e^{-\frac{\beta}{\gamma}\sum _{i=0}^n{z_i(t)}}=\nonumber\\
&&N_1+N_2+N_3-\sum_{i=0}^{\infty}{z_i(t)}-N_1e^{(\beta/\gamma)\left(N_3-\sum_{i=0}^{\infty}{z_i(t)}\right)},
\eea
and
\be
z(t)=\sum_{i=0}^{\infty}{z_i(t)},
\ee
respectively.

\subsection{Comparison with the exact numerical solution}

In order to test the applicability of the power series solutions solving the SIR model we will compare the truncated Laplace-Adomian series with the exact numerical solution. To perform the comparison we will consider only the first three terms in the expansion of $z$, so that $z(t)\approx z_0(t)+z_1(t)+z_2(t)$, or, explicitly,
\bea
z(t)&\approx& N-\left(N_1+N_2\right) e^{-\gamma t}+\frac{\gamma  N_1 \left[e^{-\beta  \left(N_1+N_2\right)t}-e^{-\gamma
   t}\right]}{\beta  \left(N_1+N_2\right)-\gamma }+\gamma  N_1^2 e^{-\left[\gamma +2 \beta  \left(N_1+N_2\right)\right]t}\times \nonumber\\
  && \frac{ \left\{-\beta
   \left(N_1+N_2\right) e^{\gamma  t}+\left[2 \beta  \left(N_1+N_2\right)-\gamma
   \right] e^{\beta   \left(N_1+N_2\right)t}+\left[\gamma -\beta  \left(N_1+N_2\right)\right] e^{2 \beta
    \left(N_1+N_2\right)t}\right\}}{\left(N_1+N_2\right) \left[\beta
   \left(N_1+N_2\right)-\gamma \right] \left[2 \beta  \left(N_1+N_2\right)-\gamma \right]}.
\eea

\begin{figure*}[tbp]
 \centering
 \includegraphics[scale=0.65]{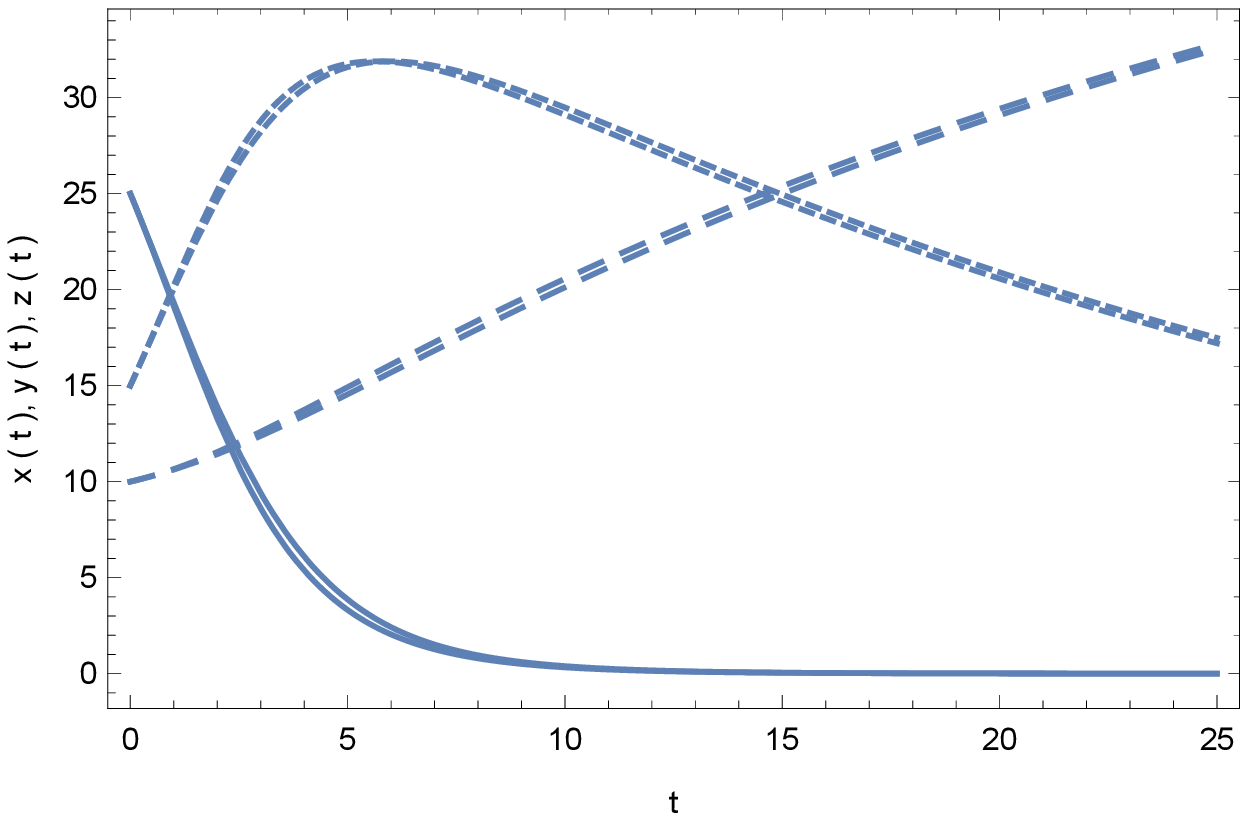}
 \includegraphics[scale=0.65]{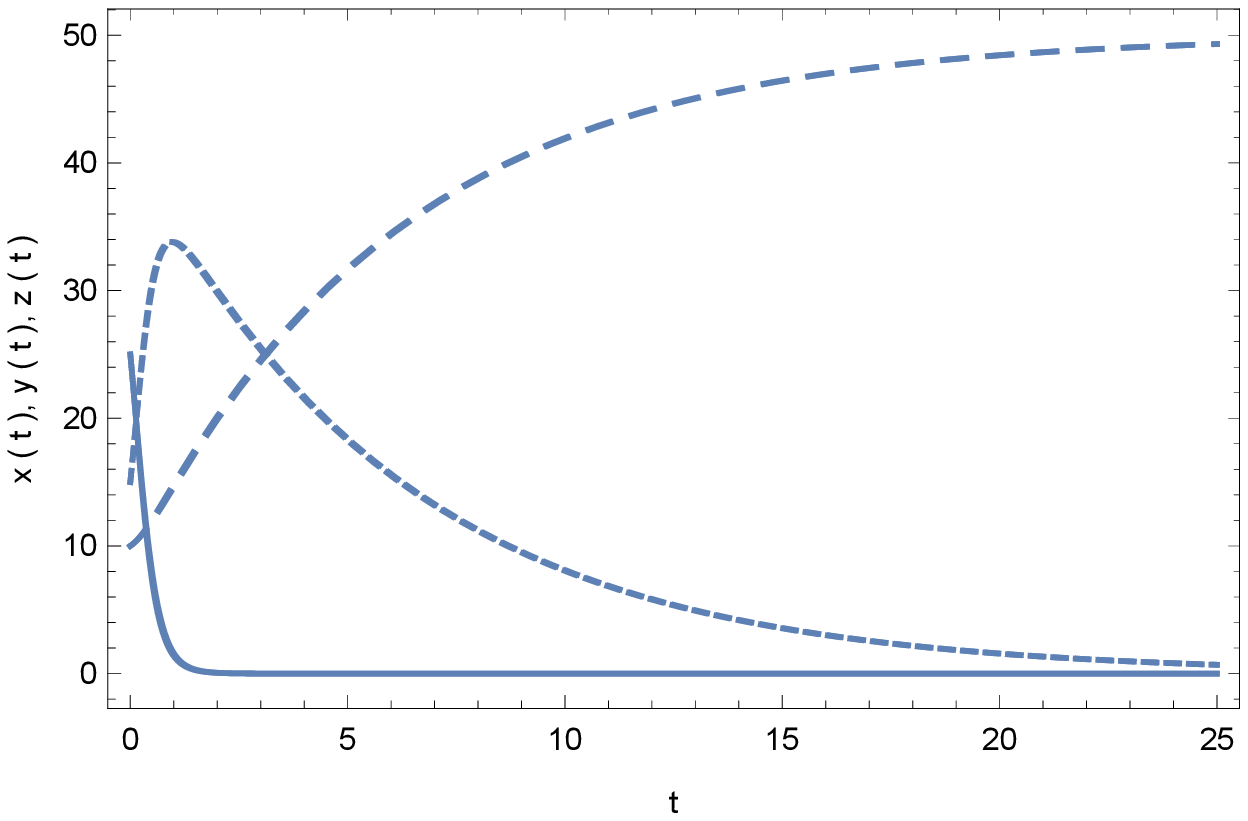}
 \caption{Comparison of the exact numerical solution of the SIR model and of the Laplace-Adomian series expansion truncated to three terms, for $\beta =0.0147$ and $\gamma =0.0358$ (left figure), and $\beta =0.098$ and $\gamma =0.164$ (right figure). $x(t)$ is represented by the solid curve, $y(t)$ by the dotted curve, and $z(t)$ by the dashed curve. The initial conditions of the SIR system are $x(0)=25$, $y(0)=15$, and $z(0)=10$.}
 \label{fig1}
\end{figure*}

\begin{figure*}[tbp]
 \centering
 \includegraphics[scale=0.65]{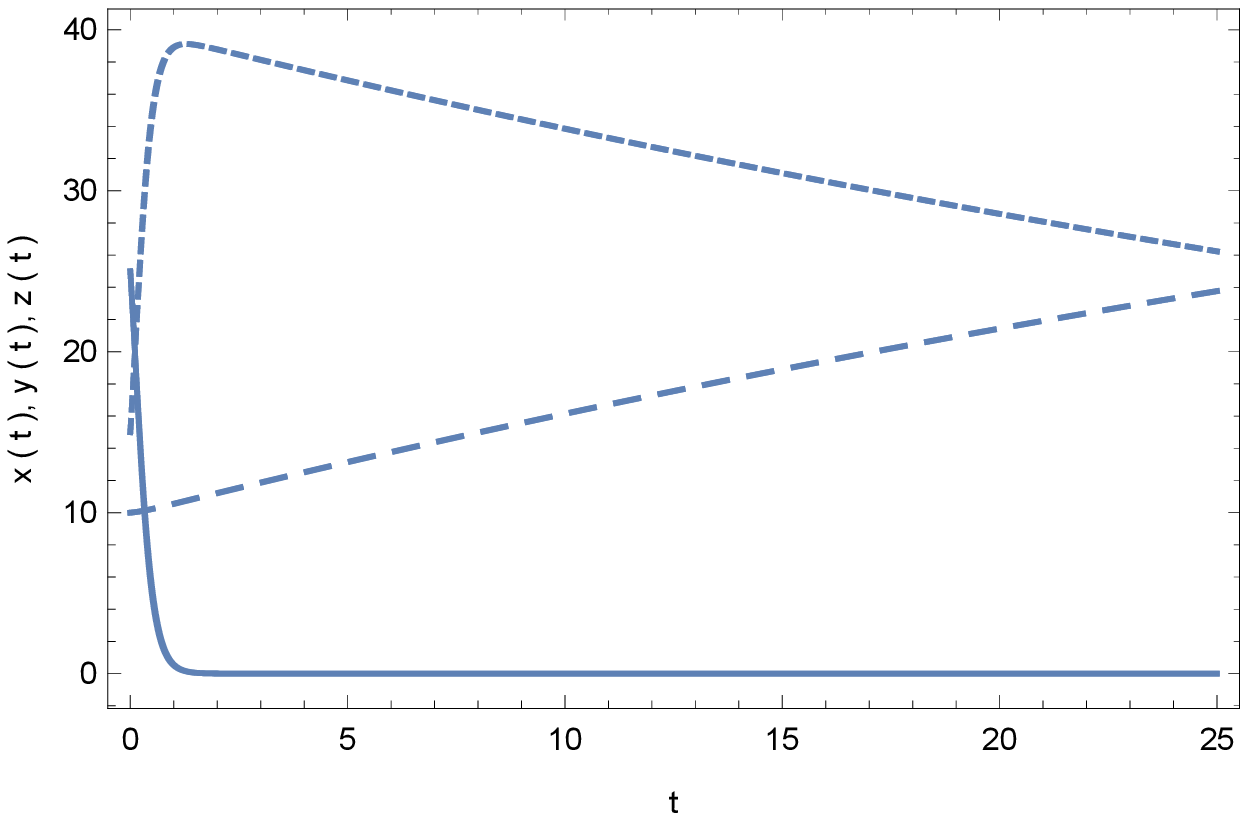}
  \includegraphics[scale=0.65]{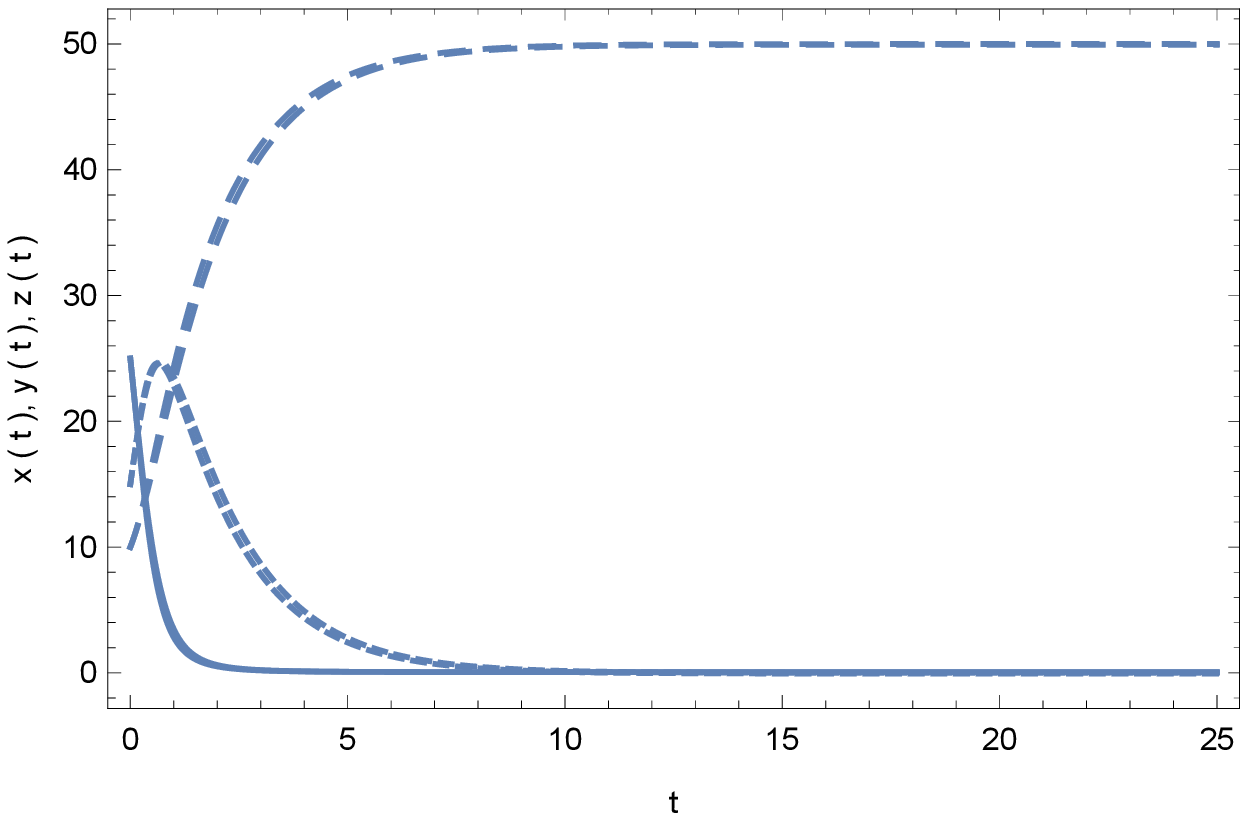}
 \caption{Comparison of the exact numerical solution of the SIR model and of the Laplace-Adomian series expansion truncated to three terms, for $\beta =0.119$ and $\gamma =0.017$ (left figure) and $\beta =0.0887$ and $\gamma =0.584$ (right figure). $x(t)$ is represented by the solid curve, $y(t)$ by the dotted curve, and $z(t)$ by the dashed curve. The initial conditions of the SIR system are $x(0)=25$, $y(0)=15$, and $z(0)=10$.}
 \label{fig2}
\end{figure*}

We fix the initial conditions of the SIR system as $x(0)=N_1=25$, $y(0)=N_2=15$, and $z(0)=N_3=10$, and we will vary the values of the parameters $\beta $ and $\gamma$. The results of the comparison between the exact numerical solution and the truncated series solutions are represented, for four distinct sets of values of $(\beta , \gamma)$, in Figs.~\ref{fig1} and \ref{fig2}, respectively.

As one can see from the Figures, the three terms truncated Laplace-Adomian series gives a very good description of the time dynamics and evolution of the SIR epidemiological model. If necessary, the precision of the semi-analytical solution can be easily increased by adding more terms in the series expansion.

\section{Conclusions}\label{IV}

Even that the exact solution of the SIR model is known, a simple but highly accurate representation of the model solution would significantly simplify the comparison of the model predictions with the epidemiological data, also allowing an efficient determination of the model parameters $\beta $ and $\gamma$ from the data fitting. The exact solution of the SIR model is given in a parametric form by $x=x_0u$, $y=N+\left(\gamma /\beta\right)\ln u -x_0u$, $z=-\left(\gamma /\beta\right)\ln u$, and $t-t_0=\int_{u_0}^u{d\xi /\left[\xi\left(x_0\beta \ln \xi-\gamma \xi-N\right)\right]}$ \cite{Harko1}, and it requires the numerical evaluation of the time integral for each step.

In the present paper we have presented a series solution of the three compartment SIR epidemiological model, giving the explicit time dependence of $x(t)$, $y(t)$ and $z(t)$. Despite its mathematical simplicity, the semianalytical solution describes very precisely the numerical behavior of the model for arbitrary values of $\beta $ and $\gamma$. The time dependence of the compartments in the semianalytical solution is given by the sum of exponential terms, and such a representation may facilitate the fitting with the statistical results. The series truncated to three terms only already give a very good description of the numerical results for arbitrary values of $\beta $ and $\gamma$, and of the initial conditions. In fact the two terms approximation also gives a good approximation of the results of the numerical integration of the SIR model.

Exact solutions of the epidemiological models are important because they can be used by epidemiologists to study the evolution of infectious diseases in different environments, and to develop the best social strategies for their control. Hopefully the results of the present paper could also contribute to the better understanding of the dynamics of the present and of the future epidemics.


\end{document}